# The Road to Useful Quantum Computers

Timothy Proctor[1,*], Robin Blume-Kohout[2], Andrew Baczewski[3]

[1]Quantum Performance Laboratory, Sandia National Laboratories, Livermore, CA 94550, USA
[2]Quantum Performance Laboratory, Sandia National Laboratories, Albuquerque, NM 87185, USA
[3]Quantum Algorithms & Applications Collaboratory, Sandia National Laboratories, Albuquerque, NM 87185, USA
*e-mail: tjproct@sandia.gov

Building a useful quantum computer is a grand science and engineering challenge, currently pursued intensely by teams around the world. In the 1980s, Richard Feynman and Yuri Manin observed independently that computers based on quantum mechanics might enable better simulations of quantum phenomena (Feynman 1982, Manin 1980). Their vision remained an intellectual curiosity until Peter Shor published his famous quantum algorithm for integer factoring (Shor 1994), and shortly thereafter a proof that errors in quantum computations can be corrected (Shor 1995). Since 1995, quantum computing R&D has progressed rapidly, from small-scale experiments in university physics laboratories (Gulde et al. 2003, DiCarlo et al. 2009) to well-funded industrial efforts and prototypes (Bluvstein et al. 2023, Kim et al. 2023, Moses et al. 2023, Chen et al. 2024, Google Quantum AI and Collaborators 2025).

Hype notwithstanding, quantum computers have yet to solve scientifically or practically important problems—a target often called *quantum utility* (Proctor 2025). In this article, we describe the capabilities of contemporary quantum computers, compare them to the requirements of quantum utility, and illustrate how to track progress from today to utility. We highlight key science and engineering challenges on the road to quantum utility, touching on relevant aspects of our own research.

**The promise of quantum computing**

Quantum computers use the rules of quantum mechanics to implement a fundamentally different kind of computation. Although many paradigms or "models" of quantum computation have been considered, including adiabatic, analogue, and measurement-based, the approach that is widely considered most promising is circuit- or gate-based quantum computing using *qubits*.

*Qubits and quantum programs*

A qubit is a quantum system that can be in either of two states—labeled "0" and "1" —*or* in a *coherent superposition* of them (Nielsen and Chuang 2012). Qubits generalize the *bits* at the heart of conventional computers. They're a little bit like probabilistic bits—e.g., when measured, a qubit probabilistically returns 0 or 1—but they are fundamentally different from them, leveraging fundamental differences between quantum and classical physics.

Quantum computations proceed via a sequence of logical instructions—gates and/or measurements—performed on a register of qubits. These operations, which transform the joint state of the register's qubits, are the quantum equivalent of machine language or microcode. Quantum programs are typically represented by a schedule of logical instructions called a *quantum circuit* (see schematic in the inset of Fig. 1a). The last step of a quantum program is to *measure* or "read out" each qubit, which yields a string of bits that is the output of the program. Because the result of measuring a qubit is probabilistic, each execution of a quantum program can return different bits, so many executions or "shots" are typically run to gather statistics.

Quantum computers are expected to solve *some* problems faster than any conventional computer, including supercomputers. Such speedups are not achieved by executing instructions faster, but by using novel quantum algorithms that use fewer instructions than algorithms for conventional computers. The art of designing these algorithms is, in part, making clever use of quantum superposition *during* the computation, to amplify the probability of bit strings that encode a solution to the problem. Quantum speedups come from a quantum algorithm's ability to take shortcuts, through the larger space of quantum states, to transform the bits encoding a problem into the bits encoding its solution.

Qubits, and thus quantum computers, can be engineered from a wide variety of physical systems. Promising qubit technologies include neutral atoms (Bluvstein et al. 2023) or ions (Moses et al. 2023, Chen et al. 2024) trapped and controlled by lasers and electromagnetic fields, superconducting circuits controlled by microwaves (Arute et al. 2019, Kim et al. 2023), electrons or nuclei in silicon wafers controlled electronically (Mądzik et al. 2022), and photons in waveguides (Aghaee Rad et al. 2025). Each qubit technology presents unique engineering problems, but many of the biggest challenges are common to all of them. Notably, most qubits must be cryogenically cooled *and*, at the same time, precisely manipulated by control systems whose complex electronics must penetrate the cooling system. These conflicting requirements are the source of noise and errors – the *bête noire* of quantum computation.

*The challenge of noise*

Real qubits, and real quantum logic operations, are always imperfect. Qubits experience noise and errors that corrupt quantum computations. Their superposition states lose coherence over time—this is called *decoherence*—due to tiny, uncontrolled interactions with a qubit's environment. Each kind of qubit is vulnerable to different sources of decoherence, but the most common cause is stray electric and magnetic fields caused by electrical equipment or even single-atom impurities. Quantum engineers go to extraordinary lengths to protect qubits from their surroundings, but they must leave gaps in that protection because qubits need to be controlled by users. The gates and measurements that drive computations are typically implemented using laser light or microwave pulses, and finite precision in these control pulses is an unavoidable source of errors. Understanding the nature of a given quantum computer's errors is an unending quest—e.g., control errors can be systematic or stochastic, and usually vary unpredictably with environmental parameters.

Qubit noise and errors harm quantum computations in diverse, complex ways (Hashim et al. 2025, Proctor et al. 2021). But the key principles can be understood by considering the simplest kind of errors: random bit flips. A bit-flip error swaps one qubit's logical 0 and 1 states, randomly, with some probability $\varepsilon$. During a quantum computation, bit-flip errors may occur on any and all qubits at random times. Just one bit-flip error will typically ruin an entire computation, because logic operations couple the flipped qubit to other qubits, spreading the error like a plague. Unless errors are somehow caught and corrected, the computation will yield a completely random (and thus useless) output.

The rate of errors places a strict limit on the size of programs that can be run reliably. If a given program applies $d$ layers of gates to $n$ qubits, we say that its *size* is $s = n \times d$ *quops* (quantum operations). A program of size $s$ offers $s$ opportunities for a failure, so it will probably fail unless $\varepsilon \ll 1/s$. In the early years of quantum computing, state-of-the-art error rates were 1-10% (Gulde et al. 2003, DiCarlo et al. 2009), permitting only *tiny* quantum programs of 100 quops or less to be run. Such tiny quantum programs like this cannot compute anything objectively useful. But since then, R&D breakthroughs have yielded qubits with much lower error rates. Moreover—and *crucially*—errors can also be corrected, albeit at great cost. So, how close are contemporary quantum computers to achieving quantum utility?

**Contemporary quantum computing**

To understand the landscape of contemporary quantum computing we need to understand both where we are (the capabilities of current hardware) and where we want to be (what quantum utility looks like).

*Contemporary quantum computing hardware*

State of the art quantum computing processors now comprise tens to hundreds of qubits (Bluvstein et al. 2023, Kim et al. 2023, Moses et al. 2023, Chen et al. 2024, Google Quantum AI and Collaborators 2025). These quantum computers have far lower error rates and many more qubits than state-of-the-art devices just 5-10 years ago. However, it is surprisingly hard to precisely quantify their performance, because errors affect them in complex ways. While detailed *in-situ* component characterization can probe a particular system's component errors precisely, it's hard to do accurately, and extrapolating performance impacts requires specialized expertise (Hashim et al. 2025, Blume-Kohout et al. 2025). The alternative is to measure capabilities directly using *benchmarking.* Benchmarking means running quantum programs that are specially designed to reveal and measure the consequences of errors *at scale,* enabling users to quantify a quantum computer's capability in realistic scenarios and predict its performance on other programs (Proctor 2025).

Quantum computer benchmarking is a very active area of research, and a primary focus of the Quantum Performance Laboratory (QPL). We've developed *capability benchmarks* (Blume-Kohout and Young 2020, Proctor et al. 2021, Proctor et al. 2025) that measure how a quantum computer's ability to run a program correctly varies with the program's depth (# of clock cycles) and *width* (# of qubits used). This is surprisingly tricky—useful quantum computer benchmarks need to overcome a variety of technical challenges, such as the exponential cost of simulating a generic quantum computation to find the result it will return when no errors occur—but, along with others in the community, we've developed technical solutions and techniques that now enable informative and efficient quantum computer benchmarks (Proctor et al. 2025).

We used these capability benchmarks to measure the evolving capabilities of IBM's cloud-access quantum computing hardware over the last 5 years (Proctor 2021). Fig. 1a shows the capabilities of these IBM quantum computing systems, as well as an estimate of how Google's 2019 "Sycamore" system (Arute et al. 2019)—the first to outperform a conventional computer on a specially-chosen problem—would have performed on these benchmarks. In these plots, we show which quantum programs can be run with high probability of no errors (a failure rate of no more than $1/e \approx 0.37$), summarized by their *shape*.

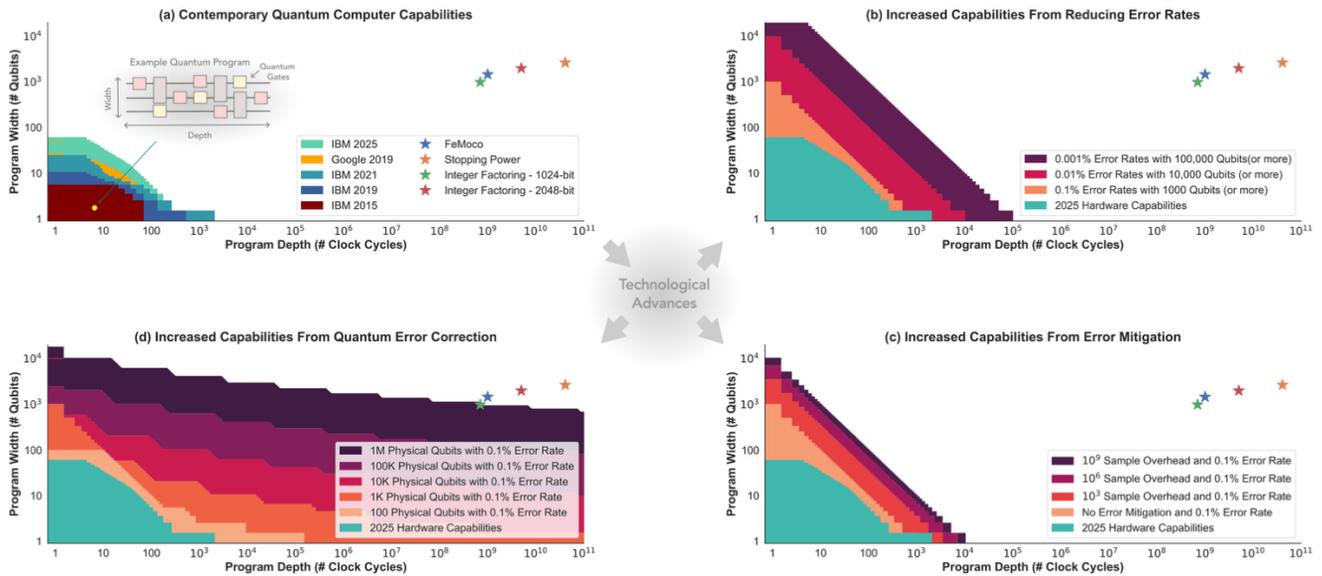

*Figure 1. There is a large gap between the capabilities of contemporary quantum computer (a, lower left) and the capabilities currently known to enable useful quantum computations (a, upper right). However, by combining three technological advances that are currently being pursued—engineering reduced error rates, error mitigation, and quantum error correction—useful quantum computing may soon be achieved. Panels (b-d) show the increase in capabilities that would be achieved with each of these advances independently, together with (large) increases in the number of (physical) qubits.*

The hardware has improved substantially over time—e.g., IBM's 156-qubit 2025-era *Kingston* device can reliably execute much larger programs than IBM's 5-qubit 2015-era *Yorktown* device, which we benchmarked in 2018. Our data probably don't even capture peak performance of contemporary IBM systems, because the performance of contemporary quantum computers varies unpredictably due to system instabilities and *Kingston* wasn't specially calibrated for our tests. Nor does this figure show capability data for contemporary trapped-ion or neutral-atom quantum computers, which have shown impressive performance in recent experiments (Bluvstein et al. 2023, Moses et al. 2023, Chen et al. 2024).

The capabilities shown in Fig. 1a are nonetheless broadly representative of what today's quantum computing hardware can do reliably. So, how close is this *status quo* to achieving quantum utility? To understand that, we need to dive a little deeper into contemporary quantum algorithms and what quantum utility is expected to look like.

*Contemporary quantum algorithms and applications*

The aim of making quantum computers more capable is ultimately to execute programs that could solve certain problems faster than any conventional computer. While quantum advantage of this sort has already been demonstrated for contrived calculations without obvious practical applications (Arute et al. 2019), the full scope of problems for which these advantages might be realized in business, science, or defense applications is still an active area of research. It is widely believed that only highly structured problems can be solved faster with quantum algorithms, and finding practical problems with such structure or discovering quantum algorithms that can exploit new structures is a grand challenge (Babbush et al*.,* 2025).

So far, the two most promising practical applications appear to be in cryptography and simulation. Public key cryptosystems (e.g. RSA) rely on the infeasibility of factoring large (e.g. 2048-bit) semiprime numbers, even on the world's biggest supercomputers. But the latest detailed analyses suggest that a quantum computer with fewer than 1 million noisy qubits could do this in a week with Shor's algorithm (Gidney 2025). Per Feynman and Manin's predictions, quantum computers are also expected to achieve higher accuracy than classical computers in simulating physical systems governed by quantum mechanics. This could have wide-ranging impacts, from the applied sciences in which it might facilitate understanding of complex molecules and materials (Bauer et al, 2020), or even fusion plasmas (Rubin et al., 2024), to the fundamental sciences in which it might help us better understand interactions between elementary particles (Bauer et al., 2023). In fact, it is becoming increasingly clear that the first useful applications of quantum computers will be physics simulations that enable scientific discovery.

While these are perhaps the best-developed algorithms and applications to date, there are still other promising quantum algorithms that could solve problems in scientific computing (e.g., studying partial differential equations or numerical linear algebra) that would be impactful in engineering applications, as well as quantum algorithms for optimization (Dalzell et al., 2025). Basic research in applications beyond simulation could be the key to quantum computers becoming more than tools for scientific discovery.

For any application, the accompanying quantum algorithm needs to be translated into a quantum computer program comprised of elementary logical operations. This program can be succinctly represented in terms of its depth and width. One can then assess whether a given quantum computer is reliable enough to execute that program for interesting instances, with a low probability of failure. For some of the best-studied problems in cryptography and simulation, Fig. 1a illustrates the gap between the capabilities of existing quantum computers and estimates for the quantum resources required to solve challenging instances of these problems. We will discuss how this gap might be bridged in the rest of this article.

**Bridging the gap to utility-scale quantum computing**

Fig. 1a shows that a vast gulf separates the capabilities of contemporary hardware from the capabilities required to run known implementations of utility-scale algorithms. Nonetheless, quantum computing technology may still achieve quantum utility as soon as the 2030s.

*Reducing noise in quantum computing hardware*

Achieving quantum utility will almost certainly require further reductions in qubit error rates *and* dramatic increases in the number of integrated qubits. These two goals have been the relentless focus of experimental quantum computing R&D since the earliest quantum computing experiments. Reducing errors can be enabled by studying the errors that are occurring in a prototype quantum computer, using *in-situ* characterization methods (Hashim et al. 2025, Blume-Kohout et al. 2025). This is a primary focus of our research at the QPL. *In situ* characterization often reveals unexpected physics or engineering problems, whose mitigation improves performance. As quantum computing hardware improves, new technical challenges emerge and must be surmounted to continue scaling up qubit count while reducing error rates. Increasingly, these are engineering challenges related to integration, qubit fabrication, and control.

There are realistic ambitions to drive all qubit error rates down to around 0.01% in the leading qubit technologies (in many technologies, the operations with the largest errors are qubit measurements and/or two-qubit gates). Some operations in some systems have already hit this milestone, but no system has yet shown capabilities consistent with these error rates. Fig. 1b shows the capability of a hypothetical "10 kiloquop" system with 0.01% error rates (but without error mitigation or correction)—as well as the capabilities of hypothetical systems with 0.1% and 0.001% error rates, corresponding to a system that is a little better than today's best systems and an *extremely optimistic* hypothetical system, respectively. This 10 kiloquop quantum computer significantly exceeds the capability of any contemporary system… and yet 10 kiloquops is *vastly* below the teraquop ($10^{12}$ quops) demands (using current best-in-class algorithm implementations) of utility-scale challenge problems. If the only way to achieve quantum utility was to reduce physical error rates, they would probably need to be reduced by another *eight orders of magnitude*! This is believed to be infeasible in any known (or imagined) qubit technology.

*Reducing the impact of noise with error mitigation*

Errors corrupt quantum computations, but if we know the error processes—or can estimate them from data—it is possible to extrapolate the corrupted result to obtain the correct, error-free result. This is known as *error mitigation* (Cai et al. 2023) and it has been used to run larger quantum computations than would have otherwise been possible (Kim et al. 2023). But error mitigation isn't free. As the amount of "noise" (the probability of an error in the program) increases, more data is required to extract the "signal" (the result that would have been obtained with no errors). A precise estimate of a program's noise-free results from error-corrupted data requires more samples—and the cost grows *exponentially* in the size of the program (Tsubouchi et al. 2023). In essence, error mitigation reduces bias in exchange for increased variance.

Fig. 1c shows an (optimistic) estimate of how error mitigation could increase the capability of our hypothetical system with 0.1% error rates—the hypothetical system that is similar to today's hardware—at the cost of three different sampling (or time) overheads. While error mitigation does increase its capability meaningfully—into the regime where simulating the quantum program with conventional computers is hard (and probably infeasible)—this hypothetical system is still nowhere near achieving gigaquop or teraquop capabilities. It is far from clear that any *useful* computations can be done in this regime. While error mitigation has a role to play, it is unlikely to enable quantum utility without the turbocharging power of *quantum error correction*.

*Removing errors with quantum error correction*

Quantum error correction is the primary ingredient needed to unlock useful quantum computing. The initial discovery that quantum error correction was possible (Shor 1995) played a central role in the emergence of quantum computing as a fully-fledged research field and potential technology in the late 1990s. Quantum error correction redundantly encodes "logical" qubits inside many physical qubits, to enable the detection and correction of errors (Brun 2020, Campbell et al. 2017). QEC has much in common with error correction for conventional computing, but it has some fascinating additional complexity. This is because the quantum error correction process must not destroy the quantum superposition in the logical qubit it protects. This prohibits, e.g., the use of a simple repetition code.

Quantum error correction enables the creation of logical qubits with much lower error rates than their constituent physical qubits. As the ratio of physical qubits to logical qubits grows, errors are suppressed *exponentially.* This enables huge increases in computational capabilities. This is shown in Fig. 1d, where we plot *approximate* capabilities of some hypothetical quantum computers with 0.1% physical-qubit error rates that use the best-known quantum error correction scheme, *surface codes* (Dennis et al. 2002). Each capability region shows the capability of a hypothetical surface-code quantum computer built from a different number of physical qubits. A device with 1 million physical qubits would be able to run the 10,000-qubit teraquop programs that would achieve quantum utility (although we note that these calculations for both capabilities and challenge problem programs are *estimates* that hide important details and include many approximations).

Quantum error correction enables and requires *fault-tolerant quantum computing* (FTQC) (Campbell et al. 2017), a quantum computing architecture (and programming model) that is radically different from the "NISQ" approach of performing computations directly with physical qubits (Preskill 2018). FTQC architectures must quickly process and act on large amounts of *syndrome* data to discover and correct errors. Furthermore, gates are more complex to implement on logical qubits, requiring weird and wonderful machinery such as *magic state distillation factories* (Litinski 2019). This has complicated and interesting consequences for compiling and scheduling algorithms in FTQC architectures (Eastin and Knill 2009, Gidney 2025, Litinski 2019).

The transition from NISQ to FTQC architectures is now well underway, and FTQC is central to the public roadmaps of many leading quantum computing companies. Since 2021, many components of FTQC have been demonstrated experimentally in multiple qubit technologies (Bluvstein et al. 2023, Google Quantum AI 2023). A landmark 2024 experiment demonstrated that QEC can indeed suppress errors (Google Quantum AI and Collaborators 2025). So, although near-term quantum utility without QEC is still being pursued (Kim et al. 2023), leading-edge research in the field is now primarily focused on FTQC architectures.

Quantum error correction only suppresses errors if the error rates of the physical qubits are low enough—below a threshold, which is a little under 1% for the surface code (Fowler et al. 2014)—and lower physical qubit error rates result in lower overheads and lower logical qubit error rates. So, by engineering FTQC systems *and* reducing the errors in the physical qubits that make up that system, capabilities can be dramatically increased. Furthermore, because logical qubits in FTQC are still not free of errors, the first useful quantum computations might combine error correction with error mitigation.

*Improving quantum algorithms*

The final piece in the puzzle of achieving quantum utility is the discovery of more efficient quantum algorithms. We have already seen that the resource requirements for problems in cryptography and simulation will require improving the quality of physical qubits such that FTQC can be implemented using quantum error correction. But because FTQC demands so much overhead, there is tremendous value in improving the quantum algorithms and their implementations so that useful problems can be solved with fewer resources—i.e., fewer logical qubits and fewer layers of instructions. Active research over the past decade has brought down resource estimates for recognized challenge problems (Reiher et al. 2017) by multiple orders of magnitude (Low et al. 2025). These breakthroughs are complemented by improvements to quantum computing architectures, which have reduced the overhead cost of FTQC. Codesign of quantum computing architectures around the algorithms that they are likely to implement will almost surely be critical to achieving quantum utility.

New quantum algorithms will not just make known problems easier and more feasible to solve on quantum computers. They have the potential to enable the solution of *new* problems that quantum computers are not yet known to help solve. But the process of translating an abstract quantum algorithm into a good

*implementation*—something that could be executed on a quantum computer to solve a real problem—is technically challenging, and the tools for analyzing and optimizing implementations of quantum algorithms are still primarily mathematical models. Even so, the supporting technologies for designing and optimizing programs for future FTQCs are developing rapidly (Harrigan et al. 2024) and credible new applications for future utility-scale quantum computers will surely follow.

## Acknowledgements

This material is based upon work supported by the U.S. Department of Energy, Office of Science, National Quantum Information Science Research Centers, Quantum Systems Accelerator (Award No. DE-SCL0000121). Additional support is acknowledged from the National Nuclear Security Administration's Advanced Simulation and Computing Program. T.P. acknowledges support from an Office of Advanced Scientific Computing Research Early Career Award. Sandia National Laboratories is a multi-program laboratory managed and operated by National Technology and Engineering Solutions of Sandia, LLC., a wholly owned subsidiary of Honeywell International, Inc., for the U.S. Department of Energy's National Nuclear Security Administration under contract DE-NA-0003525. All statements of fact, opinion or conclusions contained herein are those of the authors and should not be construed as representing the official views or policies of the U.S. Department of Energy, or the U.S. Government.